\documentstyle[12pt]{article}
\begin{document}
\noindent{\bf 2002 Oliver E. Buckley Prize. The discoveries and 
the priorities,}
\vskip0.25cm
\noindent Keshav N. Shrivastava\\
School of Physics, University of Hyderabad, Hyderabad 500046, 
India.

\vskip0.5cm
\noindent The discovery of the series of minima in the transverse 
resistivity in the quantum Hall effect as distinguished from the 
integer quantum Hall effect which gives the value of $e^2/h$ and 
the fractional value, $(1/3)e^2/h$, deserves recognition by the 
American Physical Society. The priorities in performing the 
experimental work as well as in theoretical understanding of the 
series of minima in the quantum Hall effect are pointed out. It 
is found that the sequence in which the discoveries are made, as 
recorded by the APS, is incorrect. Similarly, the assignment of 
credits for the discovery of the series, by the APS is found to 
be incorrect. Therefore, the discovery well deserves the award 
but others could have been given.

\section{Introduction}

In 1980 von Klitzing identified that $e^2/h$ can be measured from 
the plateau in the Hall resistivity. The values measured by this 
method are found to be correct. Since accurate measurement of the 
fundamental constants can be of interest, von Klitzing was 
awarded the Nobel prize by the Royal Swedish Academy of Sciences in 
1985. In 1982, Tsui, Stomer and Gossard found that the plateau 
occurs at $(1/3)e^2/h$ in addition to the one at $e^2/h$ and the 
physics of the problem giving rise to fractional effective 
charge, is different from that of the integer, $e^2/h$. 
Therefore, another Nobel prize was awarded to Tsui and Stormer in 
1998. In the first publication, it was thought that the ``fine 
structure'' is important and the numerical values of the electron 
charge, Planck's constant and that of the velocity of light were 
correct. However, later on, it was realized that the problem has 
nothing to do with the atomic fine structure. Then what the value 
of $e^2/h$ is due to? The experiment of Tsui and Stormer 
showing $(1/3)e$ is all the more puzzling but Robort Laughlin 
wrote down the wave function for a quasiparticle of charge 
$(1/3)e$ so he shared half of the Nobel prize of 1998. No one 
tried to explain the origin of either $e^2/h$ or $(1/3)e^2/h$. If 
we know the wave function of quasiparticles of charge $(1/3)e$, 
does it mean that we can explain the Hall effect? Usually, the 
Hall effect gives the concentration of carriers but the wave 
function is not used to find the Hall resistivity. So it does not 
help to know the wave function and the experiment remains 
unexplained.

In 1987, a publication appeared by Willett, Eisensein, Stormer, 
Tsui, Gossard and English [1]. Let us take only the first author 
of this paper, Willett, but what is more important is that 
earlier to this paper, the fractions reported were isolated. The 
authors wanted to report a new fraction every time, a new plateau 
was found. So there were lots of fractions in the literature. The 
paper of Willett et al changed the discovery from $e^2/h$ to a 
full series of fractions. Thus the credit of the discovery of the 
series is assigned to Willett. The experiments are done at the 
temperature of a few mK so that only a few laboratories in the 
world can do. Therefore, most likely, the assignment of the 
discovery of series of fractions to Willett is correct. In 1999, 
Willett, West and Pfeiffer showed [2] that the experimental data 
is symmetric about $\nu=1/2$. Again, taking only the first 
author, we select Willett.

The OEB prize was awarded to B.I. Halperin in 1982 and to P. A.
 Lee in 1991. Therefore, we can safely ignore Halperin and Lee from 
the credits to be given to Halperin, Lee and Read. However, 
Halperin, Lee and Read [3] is an extension of Jain [4]. Halperin 
et al used the series $\nu=p/(2p+1)$ which is the same as one of 
Jain's. Therefore, there is no need to expand on Read any further.

The motion of an electron in a curved path produces a magnetic 
field normal to the plane of the path. This is, of course, well 
known principle of making a magnet. What Jain has said is that 
even number of flux quanta, $\phi_o=hc/e$, are attached to the 
electron so that the magnetic field produced by electron with 
flux quanta attached is $B\pm 2n\phi_o$ where $n$ is the number 
of electrons per unit area. The sign depends on the alignment of 
flux quanta with respect to the external magnetic field. 
Apparently, this kind of flux attachment gives the correct series 
of fractions which are the same as those experimentally found by 
Willett.

Thus we have selected Jain, Read and Willett (in alphabetical 
order) to receive the prestigious Oliver E. Buckley (OEB) prize 
of the American Physical Society in March 2002.

\section{The Discovery.}

The value of transverse and longitudinal resistivities, 
$\rho_{xx}$ and $\rho_{xy}$ in the Hall effect of a single 
interface of GaAs/AlGaAs have been measured by Willett et al at a 
temperature of 150 mK and at high fields at 85 mK. One of the 
very good measurements is shown in Fig.1. This graph shows the 
fraction 2/5, 3/7, 5/11 and 6/13 symmetrically located on the 
right hand side of 1/2. The values on the left hand side are 2/3, 
3/5, 4/7, 5/9, 6/11 and 7/13. The resistivity is written as
\begin{equation}
\rho_{xy} = {h\over \nu e^2} 
\end{equation}
and $\nu$ determines the fractions as given above. It can be 
interpreted as the fractional charge,
\begin{equation}
e_{eff} = \nu e\,\,\,.
\end{equation}
The symmetry around $\nu=1/2$ has been emphasized again in a 
later work where it is pointed out that Jain's formula is correct.

Jain suggested that flux-quanta attached to the electron explains 
the quantum Hall effect. It produces the series of charges, 
$\nu=p/(2p\pm1)$ and then the magnetic field becomes 
$B^*=B-2n\phi_o$. Jain's series of effective charges is correct 
and agrees with the experimental data of Willett. The effective 
magnetic field is a result of having the correct series of 
charges. The expression for the effective charge is constructed 
from even number plus or minus one, with a number in the 
numerator. This construction leads to an effective magnetic 
field. The charges are, 1,2, or 3 divided by an odd number, and 
this is what is found in the experimental data. Fixing the 
charges, fixes the field as $B-2n\phi_o$ and hence even number of 
flux quanta are attached to the electron. If some how, the field 
expression can be proved to be correct, then the model can be 
accepted.

Assuming that the field expression can be found to be correct in 
the {\it future}, we can justify the award of OEB to Jain, Read and 
Willett.

\section{Is the field correct?}

Let us subject the field to a few tests.
\vskip0.25cm
\noindent{\bf(a) Flux quantization.} 

Usually the flux is quantized as,
\begin{equation}
AB = n\phi_o
\end{equation}
with $n/A=n_o$, the number per unit area, $\phi_o=hc/e$ and $B$ is 
the field. So, what is the difference between flux quantization 
above and Jain's formula,
\begin{equation}
B^*=B-2n\phi_o\qquad\mbox{(Jain)}
\end{equation}
The Jain's formula for the field is inconsistent with the flux 
quantization. If, it is a new discovery, then it need be 
consistent with flux quantization. 
\vskip0.25cm 
\noindent{\bf (b) Even feature.}

The field required by the Jain's formula is $B^*=B\pm2n\phi_o$. 
If even number of flux quanta are attached, there should be 
features at 
\begin{equation}
B-2n\phi_o~,~B+2n\phi_o~,~B-4n\phi_o~,~B+4n\phi_o~~,~\mbox{etc}
\end{equation}
when a Hall resistivity is plotted against field, there should 
occur some feature at the above fields but no such feature is 
present in the experimental data.
\vskip0.25cm
\noindent{\bf  (c) Nuclear magnetic resonance.}

Let us do the NMR to measure the magnetic field. This is a well 
known method to measure the magnetic field. We take some odd 
nuclei such as $^1H$ (proton, i.e., hydrogen in water). The 
resonance occurs when,
\begin{equation}
\omega = \gamma H
\end{equation}
where $\gamma$ is the nuclear gyromagnetic ratio. We can use the 
nuclear $g$-factor, $g_N$ and the nuclear magneton, $\mu_N$ to 
write the above expression as,
\begin{equation}
g_N\mu_NH = \hbar\omega\,\,.
\end{equation}
We can change $\omega$ from a r.f. oscillator to detect the resonance 
so that if we know $H$, we can determine $g_N$ and  of course, if 
we know $g_N$, we can determine the field. This is usually a 
continuous field. Now if CF theory is correct the Ga nuclei will 
not see $H$ but they will see $B-2n\phi_o$. The NMR experiment 
near a Hall plateau has actually been performed but such fields 
with flux attached have not been found. Some of the Ga nuclei may 
see $B$ and some others see $B^*$, then NMR will be split. No 
such splitting has been seen.

\noindent{\bf (d) Electron spin resonance.}  The electron spin 
resonance occurs at the resonance frequency determined by,
\begin{equation}
g\mu_B H = h\nu\,\,\,.
\end{equation}
If such fields as $B-2n\phi_o$ are present, the electrons will 
see them as,
\begin{equation}
g\mu_B(B-2n\phi_o) = h\nu\,\,\,.
\end{equation}
Many ESR experiments have been done but such fields with flux 
attached have never been found.

\noindent{\bf (e) Biot and Savart's law.}  The field is 
proportional to the current. Therefore, additional flux can not 
be added to $B$. Therefore, the flux attachment violates the 
elementary electrodynamics. 

In view of the above five experimental data, the idea of ``flux 
attached to electrons'' should be dropped. Jain back calculated 
the field from the quantum Hall effect data and such a field is 
not found. Recent experiments performed by Spielman et al in 
Caltech require a boson so that the composite fermion (CF) model 
will not explain the data.

\section{Statistics.}

The consistency demands that Jain's quasiparticles should be 
``composite fermions''. There is no way for them to become 
``bosons'' because of the even number of flux quanta attachment. 
The odd number of flux quanta attached shall be ``composite  
bosons''. However, it has been reported that the composite 
fermions become mixtures of bosons and fermions. Therefore, model 
of ``composite fermions'' is internally inconsistent.

Therefore, CF model is incorrect. It creates a really helpless 
situation. We are left with no theory at all for the fine 
experimental work of Bell Laboratories. Let us go back to the 
wave function of Laughlin. What the wave function does is to 
create quasiparticles of fractional charge by introducing 
``incompressibility''. However, if GaAs is compressed, the charge 
leaks and the fractionally charged particles disappear.  
Therefore, Laughlin's theory is not relevant to the experimental 
data on quantum Hall effect.

\section{Special handling.}

The APS is observing special handling of manuscripts so that only 
papers arriving from a closed group of authors are published and 
others are rejected. Therefore, only new type of interpretation 
is sought for quantum Hall effect and more conventional 
interpretations are not considered. When it comes to awarding OEB
prize, the APS members living outside U.S.A. should also be considered
but that means that there are two issues, to publish the papers is one problem and to award prizes is another. If the manuscripts of only a certain group of authors are published, the prize can still go to articles published in non-APS journals. That is very strange, the best articles are published in the APS journals and yet the prize should go to articles published in non-APS journals.

\section{The correct series.}

In the section 2 above, we pointed out that the series which 
gives the effective  fractional charge, i.e. $\nu=p/2p+1$ is correct. 
Indeed, the series $p/2p+1$ is correct but it was found by 
Shrivastava [5] at least three years before Jain. It is based on 
$l$ and $s$ values so that the effective fractional charge comes 
from the modifications of the Bohr mangeton. The symmetry about 
$\nu=1/2$ found by Willett is also present in Shrivastava's 
paper. If that is the case, then why $g$-values were not 
considered by the APS? The APS authors would have liked to 
discover the correct series from the $g$ values but such values 
were very high and did not agree with the data. Therefore, they 
were ignored by most of the authors. What is the meaning of $g$ 
values? You can get $g$-values by comparing the Zeeman energy, 
$g\mu_BH.S$ with the klystron frequency which can be tuned upto 
resonance. In semiconductors there is a band gap so that when 
$g\mu_BH=gap$, a different type of $g$ value emerges. Actually, 
Shrivastava considered half the $g$ value, many values of $l$ and 
$s$, including negative $s$. The entire data from PRL and PRB was 
considered and in all cases Shrivastava's theory is found to be 
correct. It is amazing that Shrivastava has a quasiparticle of 
zero charge and spin ${1\over2}$, with a new phenomenon called 
``superresistivity''. Laughlin has a wave function of charge 
1/3. In Shrivastava's paper 1/3 comes from a  certain combination 
of $l$ and $s$. Laughlin talks about spinons of spin 1/2 and zero 
charge. In Shrivastava's formula zero charge gives infinite 
resistivity and spin 1/2, which are not due to Laughlin.

It may be added that quantization of resistivity at $h/e^2$ is 
not due to atomic fine structure, it is due to flux quantization, 
and the quantization at fractional charge is not due to 
quasiparticles of fractional charge and fluxes are not attached 
to electrons. The fractional quantization occurs due to 
combinations of $l$ and $s$.

The correct theory of the quantum Hall effect which explains the 
experimental data is given in a recently published book [6].

\section{Conclusions.}

There is no doubt that the fine experimental work of Willett is 
different from that of integer quantized Hall effect for which 
von Klitzing was awarded the Nobel prize in 1985.
The experimental identification of series of effective charges is 
also different from that of fractional plateaus found by Tsui and 
Stormer which was also awarded the Noble prize of 1998. It is 
thought that Read's papers are developments of original work done 
by others. Jain's back calculation of fields from the experimental 
data is surely not correct and the claim made that flux quanta 
are attached to the electrons is not justified and the award of 
the OEB prize relies on possible future developments. 
Shrivastava's work published several years before Jain, is 
pointed out as the correct interpretation of the quantum Hall 
effect.

\noindent\hrulefill

Keshav Shrivastava obtained his Ph.D. from the Indian Institute 
of Technology in 1966. He worked in the UC Santa Barbara, Univ. 
Houston, Univ. Nottingham, the State University, Utrecht, etc. He 
has published 170 papers in the last 36 years. He is the author 
of Superconductivity: Elementary Topics, World Scientific, 2000.

\noindent \hrulefill

\noindent{\bf References}
\begin{enumerate}
\item R. Willett, J.P. Eisenstein, H.L. Stormer, D.C. Tsui, A.C. 
Gossard, and J.H. English, Phys. Rev. Lett. {\bf59}, 1776 (1987).
\item R.L. Willett, K.W. West and L.N. Pfeiffer, Phys. Rev. lett. 
{\bf83}, 2624 (1999).
\item B.I. Halperin, P.A. Lee and N. Read, Phys. Rev. B{\bf47}, 
7312 (1993).
\item J.K. Jain, Phys. Rev. Lett. {\bf63}, 199 (1989).
\item K.N. Shrivastava, Phys. Lett. A{\bf113}, 435 (1986); 
A{\bf115}, 459 (1986)(E).
\item K.N. Shrivastava, Introduction to Quantum Hall Effect, Nova 
Science Publishers Inc., New York (2002).
\end{enumerate}
\begin{itemize}
\item[{\bf Fig.1:}] The integer quantized Hall effect and the 
fractional quantized Hall effect have disappeared and a ``series 
quantized Hall effect has appeared when data is recorded 
properly. The experimental data of Willett et al is shown. The 
series on the right hand side of 1/2 as well as that on the left 
hand side are the same as in ref. 5.
\end{itemize}

\end{document}